\documentclass{elsart}
\usepackage[dvips]{graphicx}
\usepackage{subfigure}
\usepackage[nolists,heads]{endfloat}
\usepackage{colortbl}
\definecolor{cLightYellow}{rgb}{.90,.85,.55}
\definecolor{cLightGray}{rgb}{.90,.90,.90}
\definecolor{cMediumGray}{rgb}{.70,.70,.70}
\begin{document}
\begin{frontmatter}
\title{Pareto's Law of Income Distribution: Evidence for Germany, the United Kingdom, and the United States}
\author[Roma,S.I.E.C.]{F. Clementi\corauthref{cor}},
\corauth[cor]{Corresponding author. Tel.: +39--06--49--766--843; fax: +39--06--44--61--964.}
\ead{fabio.clementi@uniroma1.it}
\author[Ancona,S.I.E.C.]{M. Gallegati}
\ead{gallegati@dea.unian.it}
\address[Roma]{Department of Public Economics, University of Rome `La Sapienza', Via del Castro Laurenziano 9, 00161 Rome, Italy}
\address[Ancona]{Department of Economics, Universit\`a Politecnica delle Marche, Piazzale Martelli 8, 60121 Ancona, Italy}
\address[S.I.E.C.]{S.I.E.C., Universit\`a Politecnica delle Marche, Piazzale Martelli 8, 60121 Ancona, Italy}
\begin{abstract}
We analyze three sets of income data: the US Panel Study of Income Dynamics (PSID), the British Household Panel Survey (BHPS), and the German Socio-Economic Panel (GSOEP). It is shown that the empirical income distribution is consistent with a two-parameter lognormal function for the low-middle income group (97\%--99\% of the population), and with a Pareto or power law function for the high income group (1\%--3\% of the population). This mixture of two qualitatively different analytical distributions seems stable over the years covered by our data sets, although their parameters significantly change in time. It is also found that the probability density of income growth rates almost has the form of an exponential function.
\end{abstract}
\end{frontmatter}
\section{Introduction}
More than a century ago, the economist Vilfredo Pareto stated in his {\em Cours d'\'Economie Politique} that there is a simple law which governs the distribution of income in all countries and at all times. Briefly, if $N$ represents all the number of income-receiving units cumulated from the top above a certain income limit $x$, and $A$ and $\alpha$ are constants, then:
\begin{equation}
N=\frac{A}{x^{\alpha}}
\label{eq:Pareto_Law}
\end{equation}
and, therefore, $log\left(N\right)=log\left(A\right)-\alpha log\left(x\right)$. In other words, if the logarithms of the number of persons in receipt of incomes above definite amounts are plotted against the logarithms of the amount of these incomes, the points so obtained will be on a straight line whose slope with the axis on which the values of $log\left(x\right)$ are given will be $\alpha$. Pareto examined the statistics of incomes in some countries and concluded that the inclination of the line with the $log\left(x\right)$ axis differed but little from 1.5.
\par
Very recently, considerable investigations with modern data in capitalist\linebreak economies have revealed that the upper tail of the income distribution (generally less than 5\% of the individuals) indeed follows the above mentioned behaviour, and the variation of the slopes both from time to time and from country to country is large enough not to be negligible. Hence, characterization and understanding of income distribution is still an open problem. The interesting problem that remains to be answered is the functional form more adequate for the majority of population not belonging to the power law part of the income distribution. Using data coming from several parts of the world, a number of recent studies debate whether the low-middle income range of the income distribution may be fitted by an exponential [1--8] or lognormal [9--13] decreasing function.\footnote{Recently, a distribution proposed by [14,15] has the form of a deformed exponential function:
\[P_{\kappa}\left(x\right)=\left(\sqrt{1+\kappa^{2}x^{2}}-\kappa x\right)^{\frac{1}{\kappa}}\]
which seems to capture well the behaviour of the income distribution at the low-middle range as well as the power law tail.}
\par
In this paper we have analyzed three data sets relating to a pool of major industrialized countries for several years in order to add some empirical investigations to the ongoing debate on income distribution. When fits are performed, a two-parameter lognormal distribution is used for the low-middle part of the distribution (97\%--99\% of the population), while the upper high-end tail (1\%--3\% of the population) is found to be consistent with a power law type distribution. Our results show that the parameters of income distribution change in time; furthermore, we find that the probability density of income growth rates almost scales as an exponential function.
\par
The structure of the paper is as follows. Section \ref{sec:TheData} describes the data used in our study. Section \ref{sec:EmpiricalFindings} presents and analyzes the shape of the income distribution (Section \ref{sec:TheShapeOfTheDistribution}) and its time development over the years covered by our data sets (Section \ref{sec:TemporalChangeOfTheDistribution}). Section \ref{sec:SummaryAndConclusions} concludes the paper.
\section{The Data}
\label{sec:TheData}
We have used income data from the US Panel Study of Income Dynamics (PSID), the British Household Panel Survey (BHPS), and the German Socio-Economic Panel (GSOEP) as released in a cross-nationally comparable format in the Cross-National Equivalent File (CNEF). The CNEF brings together multiple waves of longitudinal data from the surveys above, and therefore provides relatively long panels of information. The current release of the CNEF includes data from 1980 to 2001 for the PSID, from 1991 to 2001 for the BHPS, and from 1984 to 2002 for the GSOEP. Our data refer to the period 1980--2001 for the United States, and to the period 1991--2001 for the United Kingdom. As the eastern states of Germany were reunited with the western states of the Federal Republic of Germany in November 1990, the sample of families in the East Germany was merged with the existing data only at the beginning of the 1990s. Therefore, in order to perform analyses that represent the population of reunited Germany, we chose to refer to the subperiod 1990--2002 for the GSOEP.
\par
A key advantage of the CNEF is that it provides reliable estimates of annual income variables defined in a similar manner for all the countries that are not directly available in the original data sets.\footnote{Reference [16] offers a detailed description of the CNEF. See also the CNEF web site for details: http://www.human.cornell.edu/pam/gsoep/equivfil.cfm.} It includes pre- and post-government household income, estimates of annual labour income, assets, private and public transfers, and taxes paid at household level. In this paper, the household post-government income variable (equal to the sum of total family income from labour earnings, asset flows, private transfers, private pensions, public transfers, and social security pensions minus total household taxes) serves as the basis for all income calculations. Following a generally accepted methodology, the concept of {\em equivalent income} will serve as a substitute for personal income, which is unobservable. Equivalent income $x$ is calculated as follows. In a first step, household income $h$ is adjusted for by household type $\theta$ using an equivalence scale $e\left(\theta\right)$.\footnote{We use the so-called ``modified OECD'' equivalence scale, which is defined for each household as equal to $1+0.5\times\left(\#\mbox{adults}-1\right)+0.3\times\left(\#\mbox{children}\right)$.} This adjusted household income $x=h/e\left(\theta\right)$ is then attributed to every member of the given household, which implies that income is distributed equally within households.
\par
In the most recent release, the average sample size varies from about 7,300 households containing approximately 20,200 respondent individuals for the PSID-CNEF to 6,500 households with approximately 16,000 respondent individuals for the BHPS-CNEF; for the GSOEP-CNEF data from 1990 to 2002, we have about 7,800 households containing approximately 20,400 respondent individuals.
\par
All the variables are in current year currency; therefore, we use the consumer price indices to convert into constant figures for all the countries. The base year is 1995.
\section{Empirical Findings}
\label{sec:EmpiricalFindings}
\subsection{The Shape of the Distribution}
\label{sec:TheShapeOfTheDistribution}
\begin{figure}
\centering
\begin{center}
\mbox{
\subfigure[United States (1996)]{\includegraphics[width=0.48\textwidth]{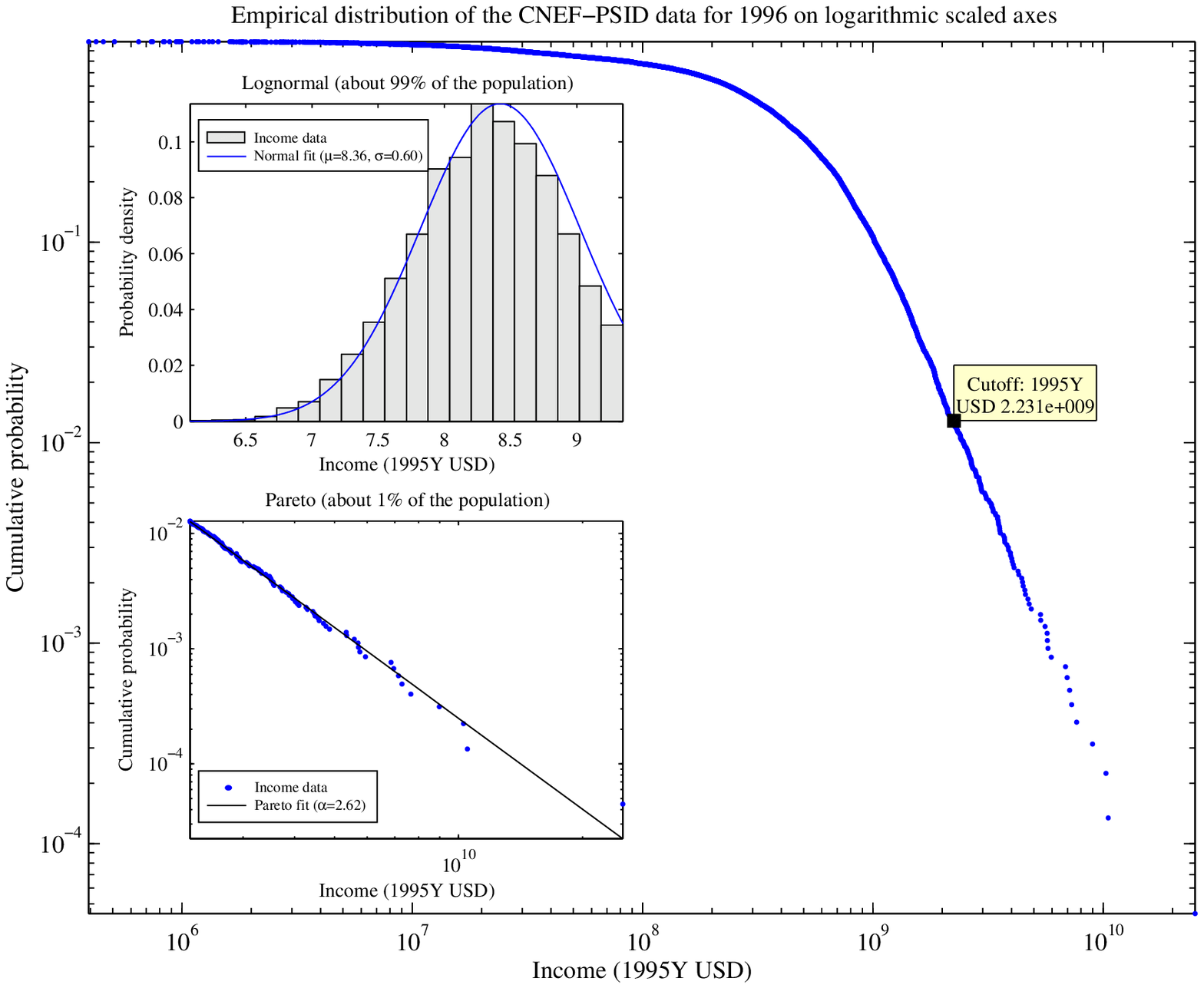}}
}
\mbox{
\subfigure[United Kingdom (2001)]{\includegraphics[width=0.48\textwidth]{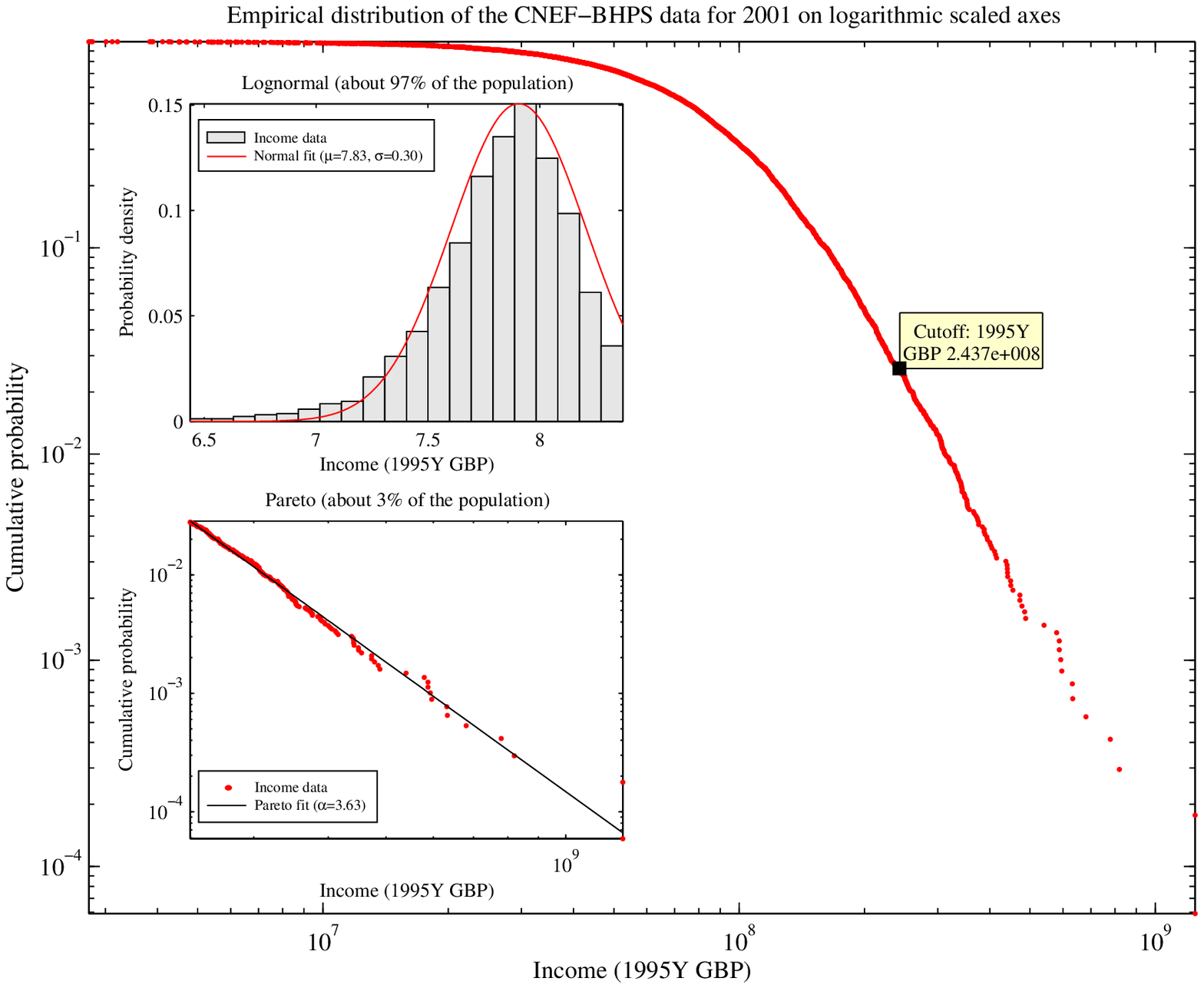}}
}
\mbox{
\subfigure[Germany (1991)]{\includegraphics[width=0.48\textwidth]{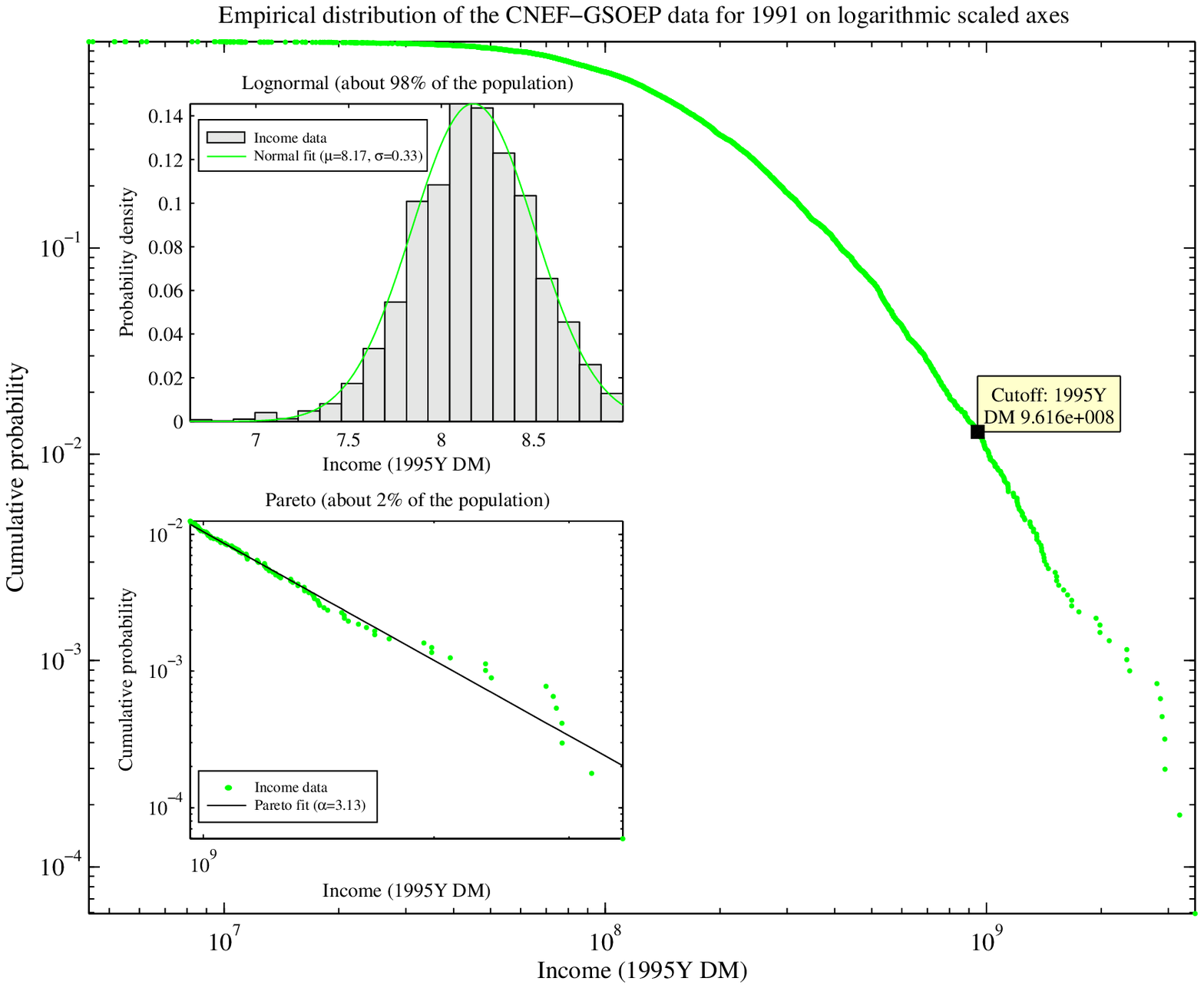}}
}
\caption{The cumulative probability distribution of the equivalent income in the log-log scale along with the lognormal (top insets) and Pareto (lower insets) fits for some randomly selected years}
\label{fig:The_Shape_of_the_Distributions}
\end{center}
\end{figure}

The main panel of the pictures illustrated in Fig. \ref{fig:The_Shape_of_the_Distributions} presents the empirical cumulative distribution of the equivalent income from our data sets for some randomly selected years in the log-log scale.\footnote{To treat each wave of the surveys at hand as a cross-section, and to obtain population-based statistics, all calculations used sample weights which compensate for unequal probabilities of selection and sample attrition. Furthermore, to eliminate the influence of outliers, the data were trimmed. We also dropped observations with zero and negative incomes from all samples.}
As shown in the lower insets, the upper income tail (about 1\%--3\% of the population) follows the Pareto's law:
\begin{equation}
1-F\left(x\right)=P\left(X\geq x\right)=C_{\alpha}x^{-\alpha}
\end{equation}
where $C_{\alpha}=k^{\alpha}$, $k,\alpha>0$, and $k\leq x<\infty$. Since the values of $x$ above some value $x_{R}$ can not be observed due to tail truncation, to fit the (logarithm of the) data for the majority of the population (until the 97\textsuperscript{th}--99\textsuperscript{th} percentiles of the income distribution) we use a right-truncated normal probability density function:
\begin{equation}
f\left(y\right)=\left\{
\begin{array}{ll}
\frac{f\left(y\right)}{\int\limits_{-\infty}^{y_{R}}f\left(y\right)dy}&,\;\;\;\;\;\;\;\;\;-\infty<y\leq y_{R}\\
0&,\;\;\;\;\;\;\;\;\;y_{R}\leq y<\infty
\label{eq:RTN_PDF}
\end{array}
\right.
\end{equation}
where $y=log\left(x\right)$, and $y_{R}=log\left(x_{R}\right)$. The fit to Equation (\ref{eq:RTN_PDF}) is shown by the top insets of the pictures.
\par
\begin{figure}
\centering
\begin{center}
\mbox{
\subfigure[United States (1996)]{\includegraphics[width=0.48\textwidth]{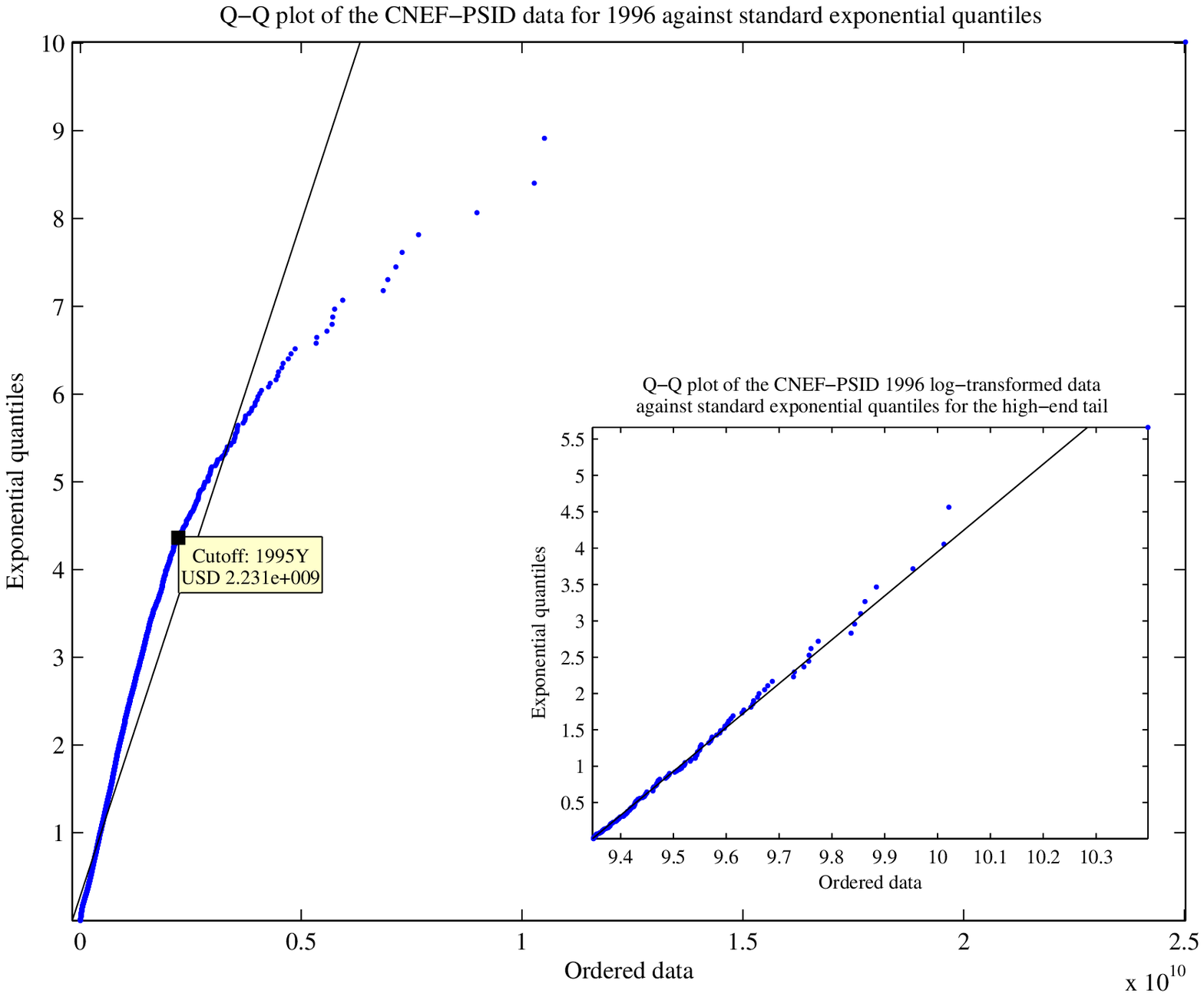}}
\subfigure[United States (1996)]{\includegraphics[width=0.48\textwidth]{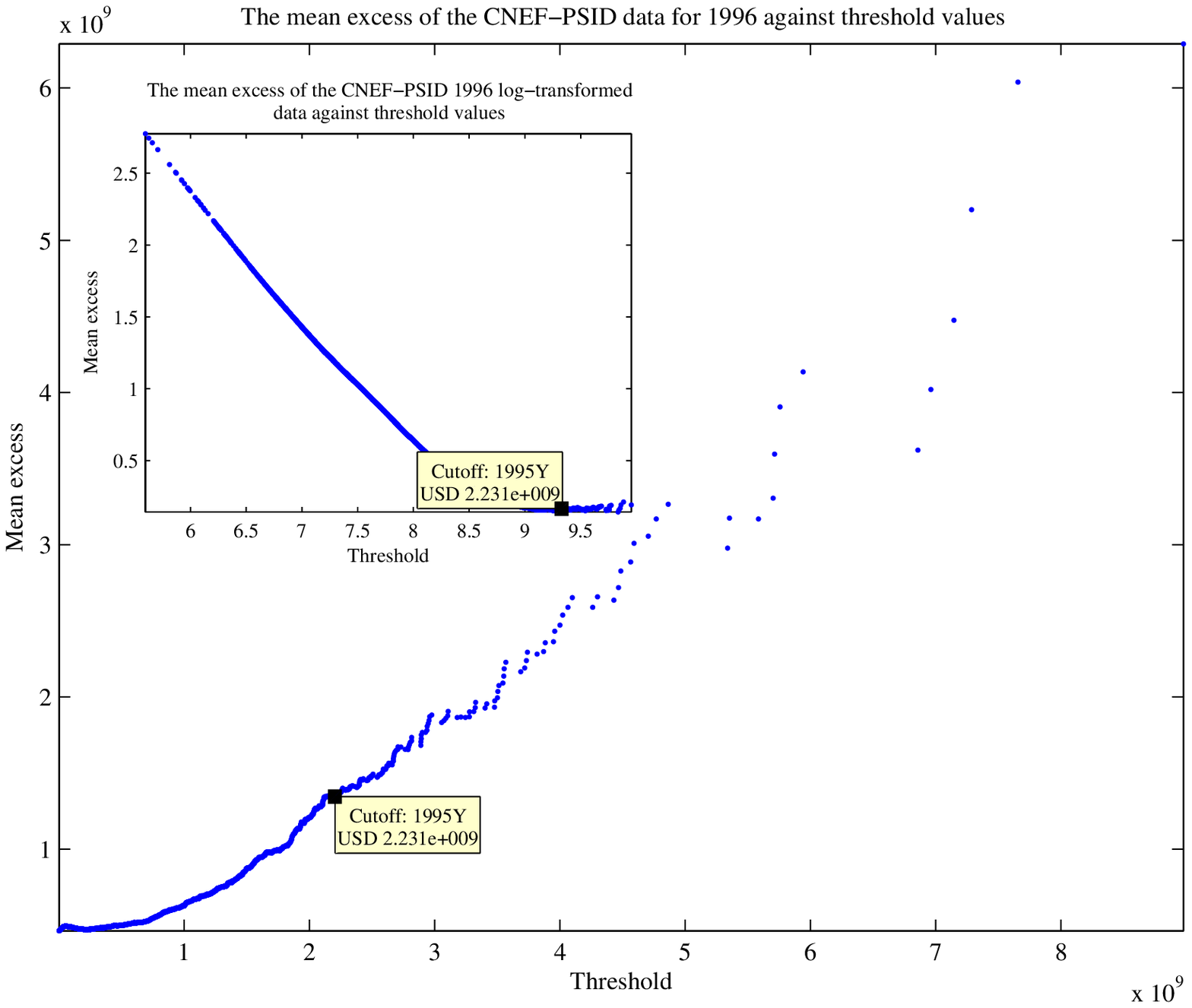}}
}
\mbox{
\subfigure[United Kingdom (2001)]{\includegraphics[width=0.48\textwidth]{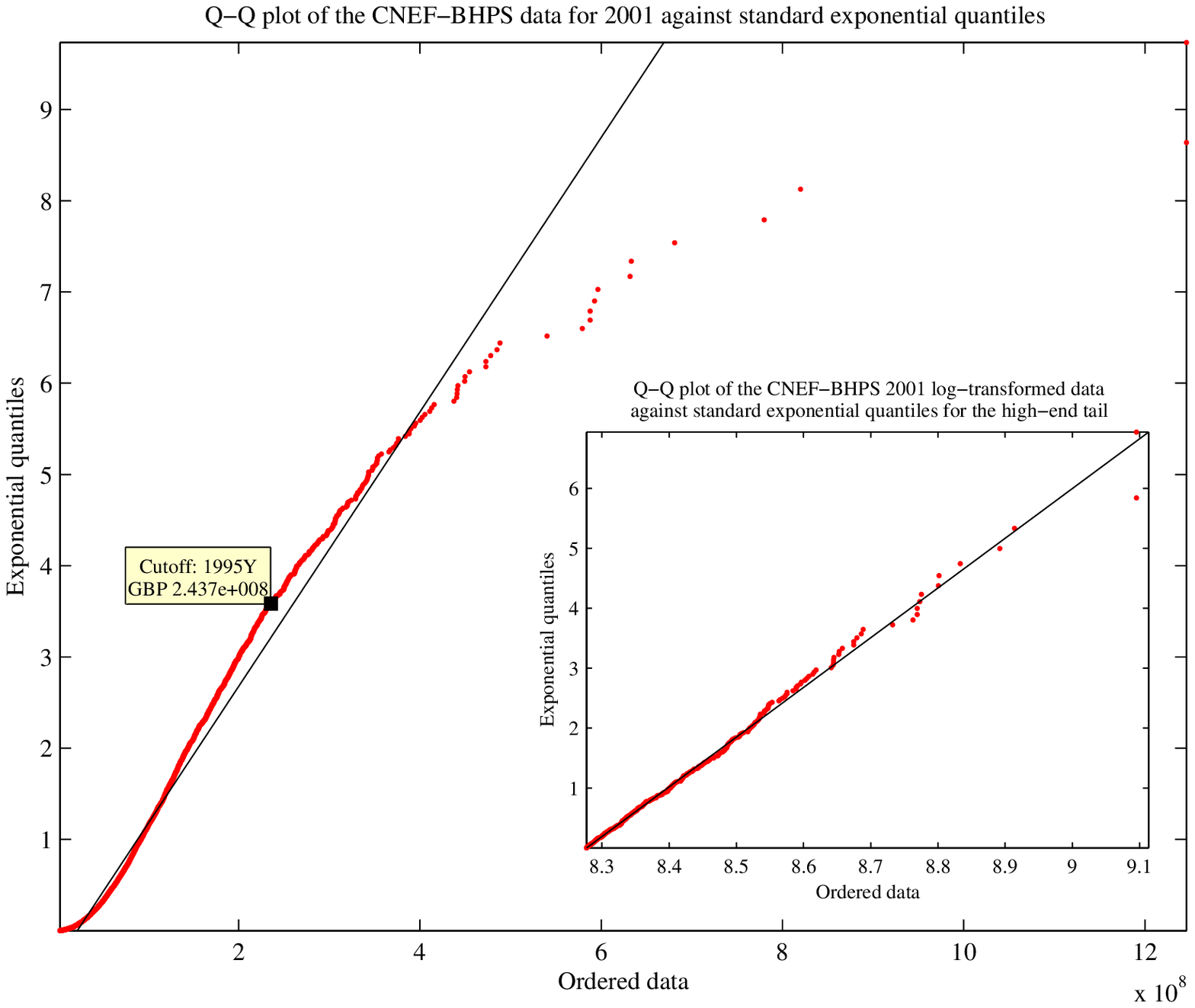}}
\subfigure[United Kingdom (2001)]{\includegraphics[width=0.48\textwidth]{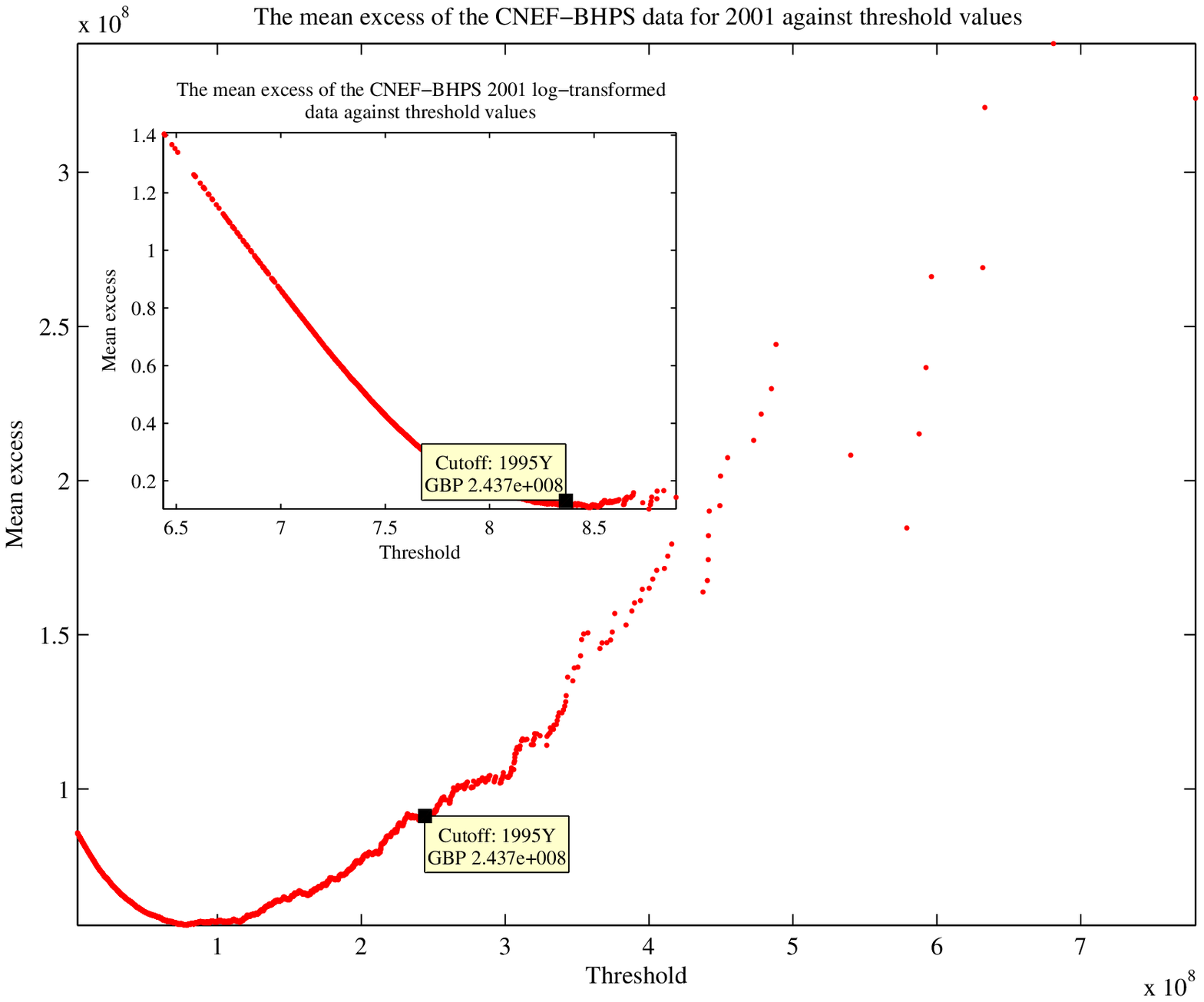}}
}
\mbox{
\subfigure[Germany (1991)]{\includegraphics[width=0.48\textwidth]{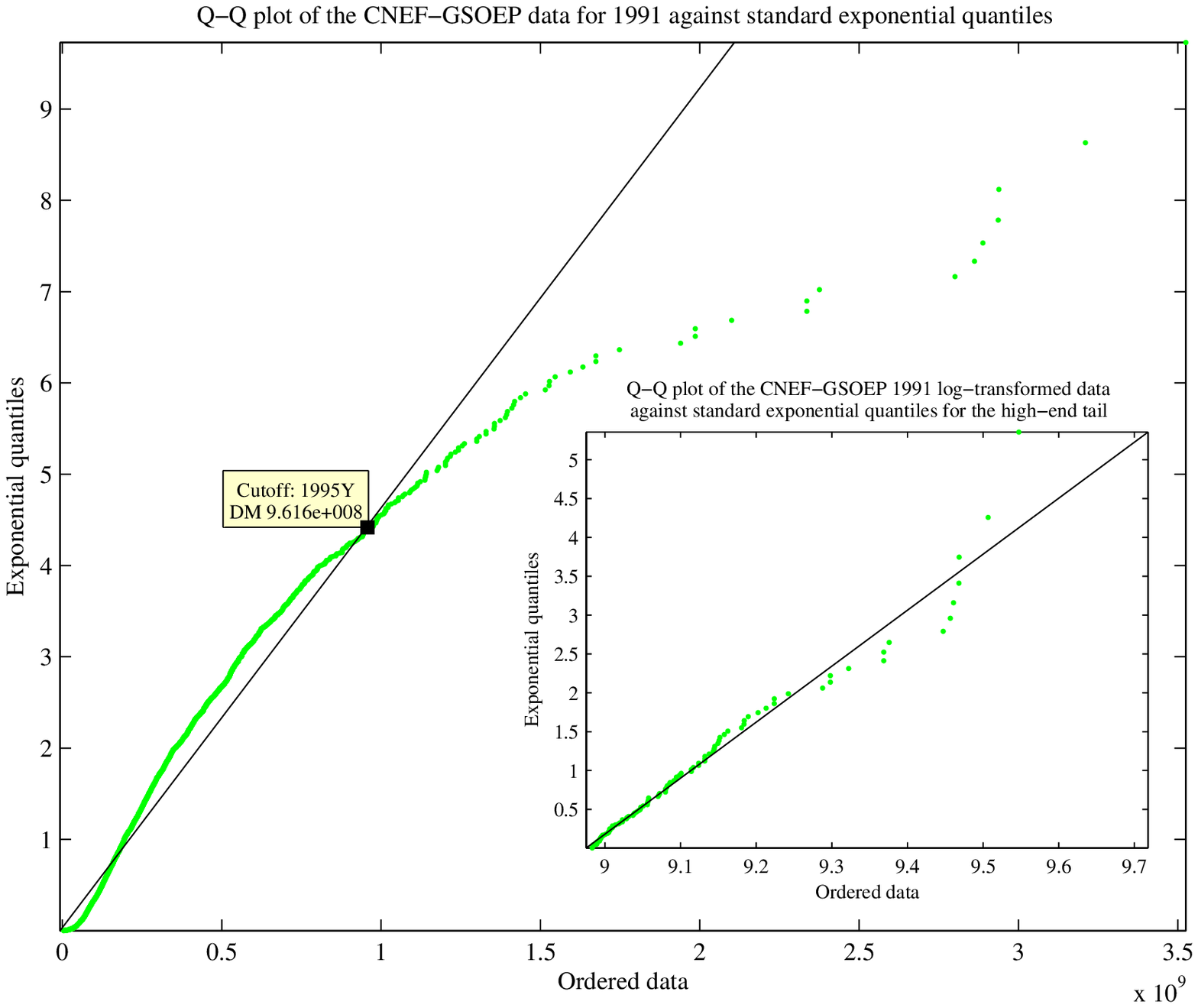}}
\subfigure[Germany (1991)]{\includegraphics[width=0.48\textwidth]{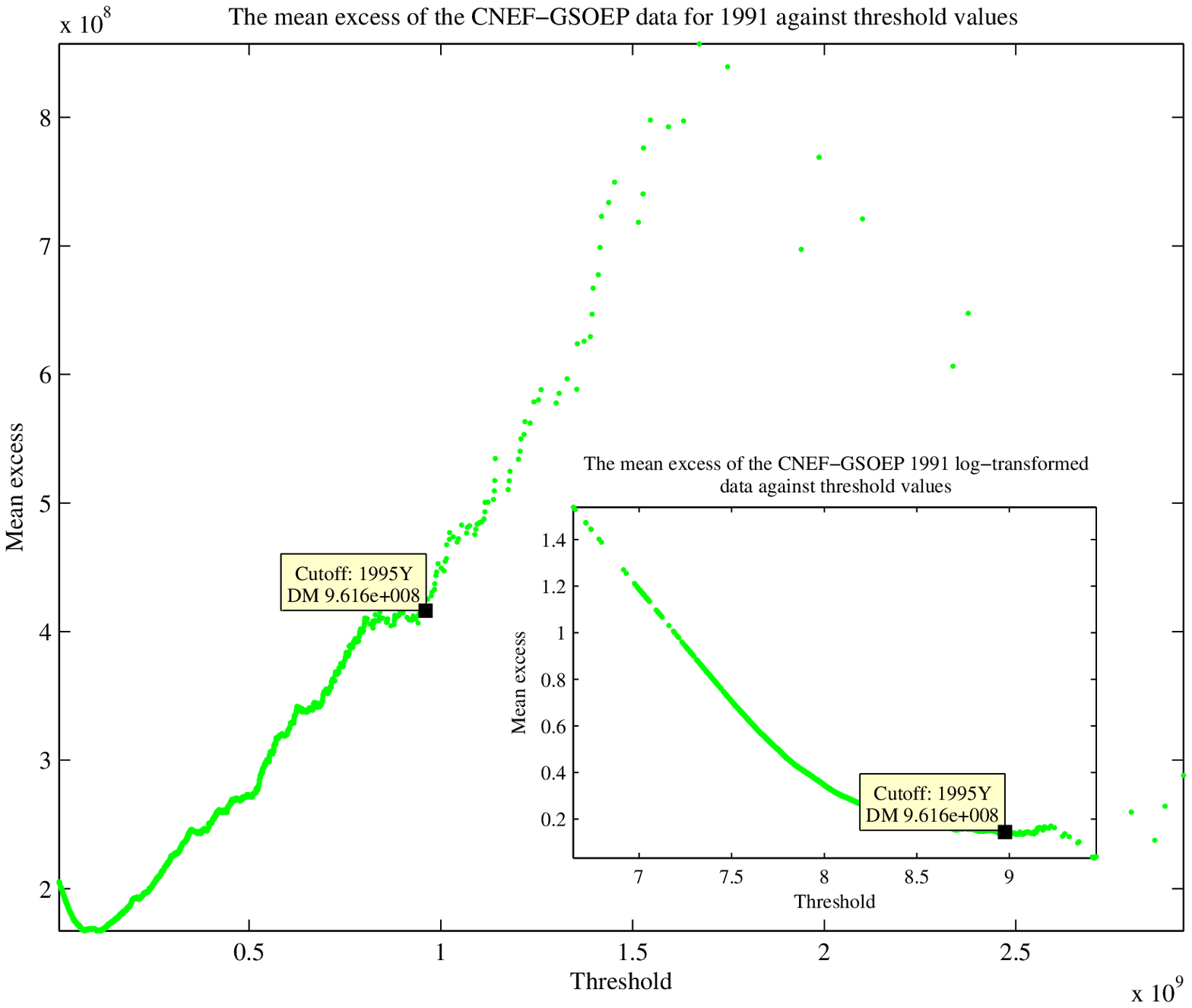}}
}
\caption{Q-Q plots (left pictures) against standard exponential quantiles and mean excess plots (right pictures) against threshold values for some randomly selected years. A concave departure from the straight line in the Q-Q plot (as in the left main panels) or an upward sloping mean excess function (as in the right main panels) indicate a heavy tail in the sample distribution. The insets in the pictures apply the same graphical tools to the log-transformed data}
\label{fig:Graphical_Tools_for_Data_Analysis}
\end{center}
\end{figure}

To select a suitable threshold or cutoff value $x_{R}$ separating the lognormal part from the Pareto power law tail of the empirical income distribution, we use visually oriented statistical techniques such as the {\em quantile-quantile} (Q-Q) and {\em mean excess} plots. Figure \ref{fig:Graphical_Tools_for_Data_Analysis} gives an example of these graphical tools for the countries at hand.
The left pictures in the figure are the plots of the quantile function for the standard exponential distribution (i.e., a distribution with a medium-sized tail) against its empirical counterpart. If the sample comes from the hypothesized distribution, or a linear transformation of it, the Q-Q plot is linear. The concave presence in the plots is an indication of a fat-tailed distribution. Since a log-transformed Pareto random variable is exponentially distributed, we conduct experimental analysis on the log-transformed data by excluding some of the lower sample points to investigate the concave departure region on the plots and obtain a fit closer to the straight line. The results are shown by the insets of the left pictures in the figure. The right pictures plot the empirical average of the data that are larger than or equal to $x_{R}$, $E\left(X|X\geq x_{R}\right)$, against $x_{R}$. If the plot is a linear curve, then it may be either a power type or an exponential type distribution. If the slope of the linear curve is greater than zero, then it suggests a power type (as in the main panels); otherwise, if the slope is equal to zero, it suggests an exponential type (as in the insets for the log-transformed data).
\subsection{Temporal Change of the Distribution}
\label{sec:TemporalChangeOfTheDistribution}
\begin{figure}
\centering
\begin{center}
\mbox{
\subfigure[United States (1980--2001)]{\includegraphics[width=0.48\textwidth]{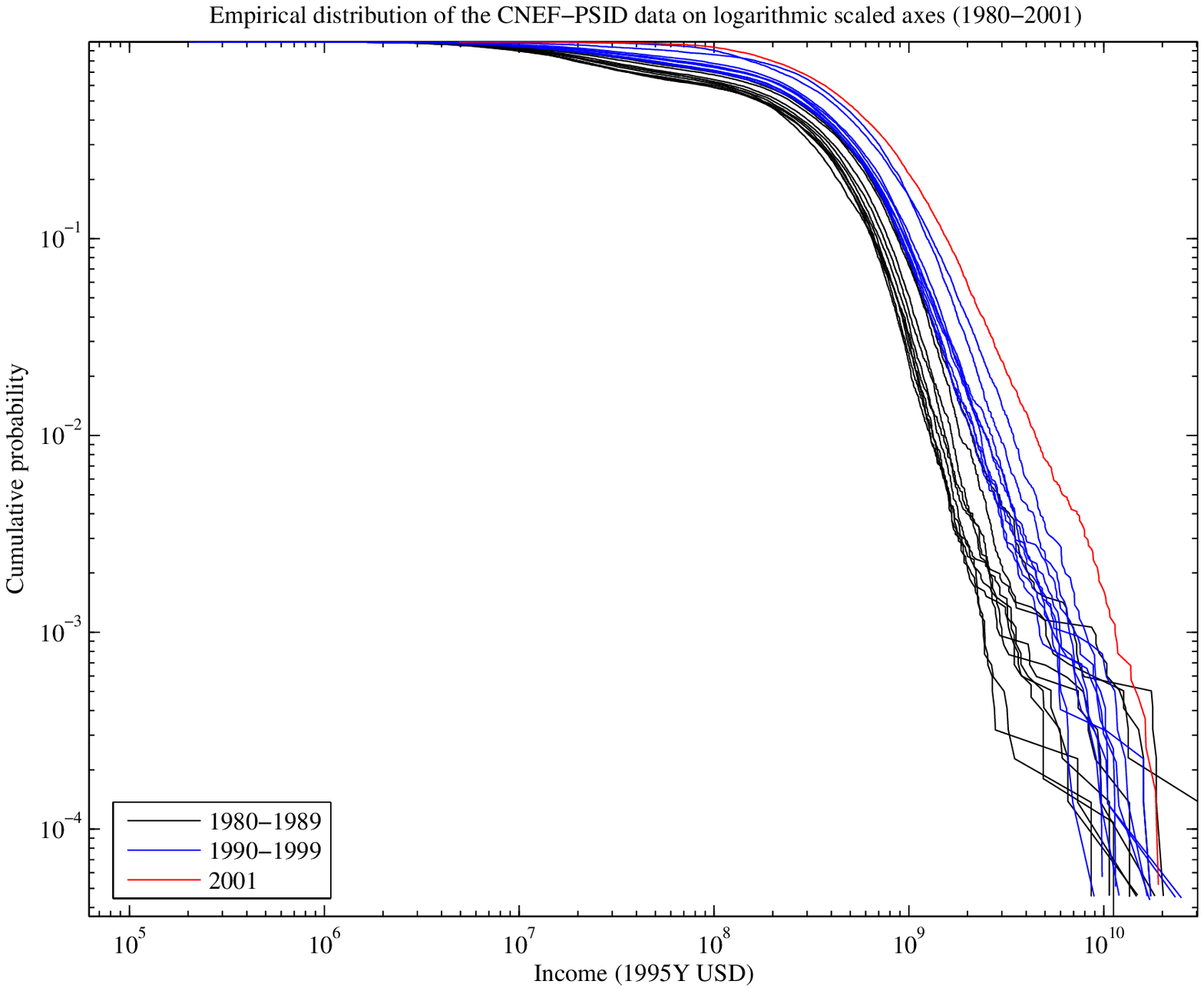}}
}
\mbox{
\subfigure[United Kingdom (1991--2001)]{\includegraphics[width=0.48\textwidth]{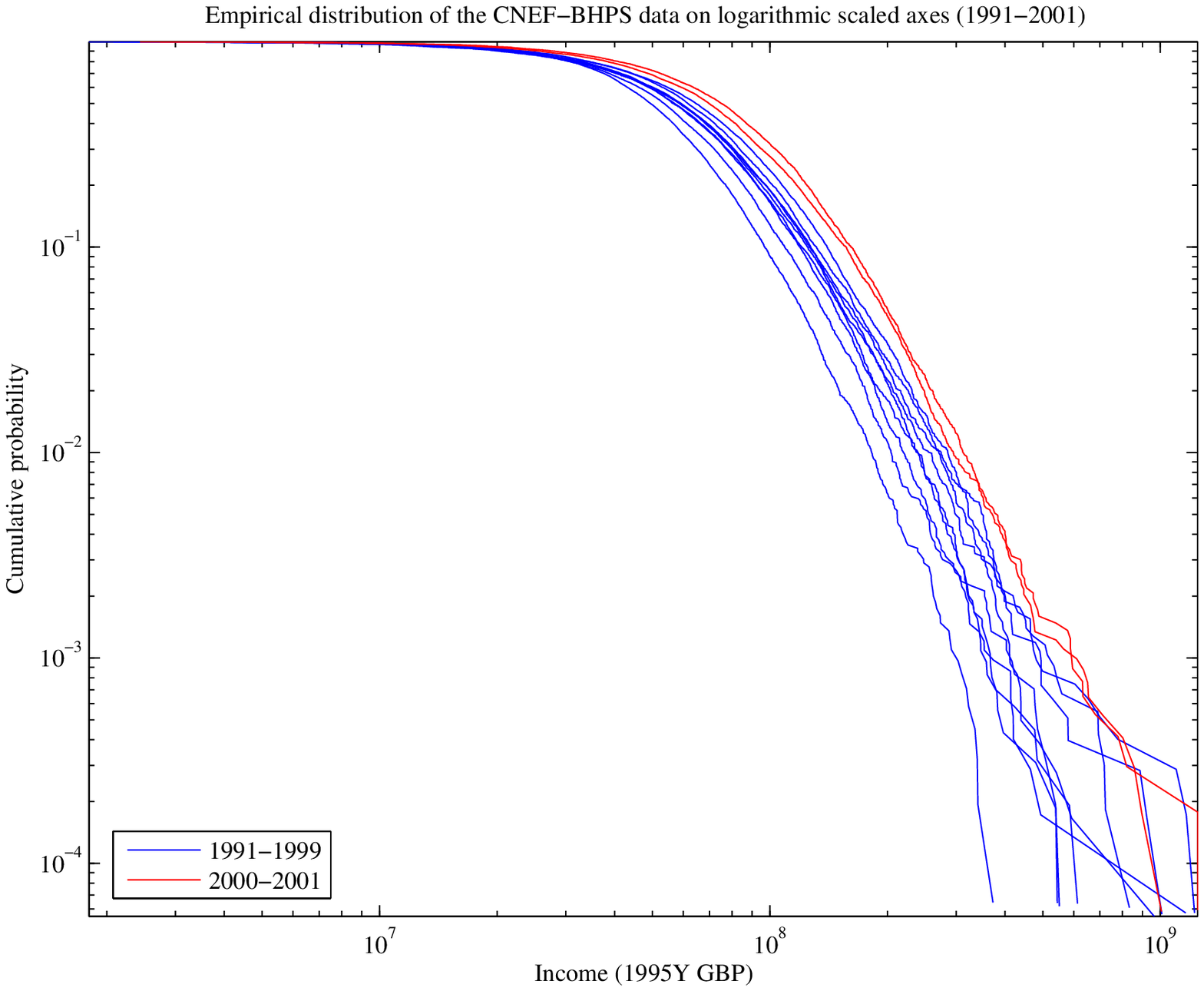}}
}
\mbox{
\subfigure[Germany (1990--2002)]{\includegraphics[width=0.48\textwidth]{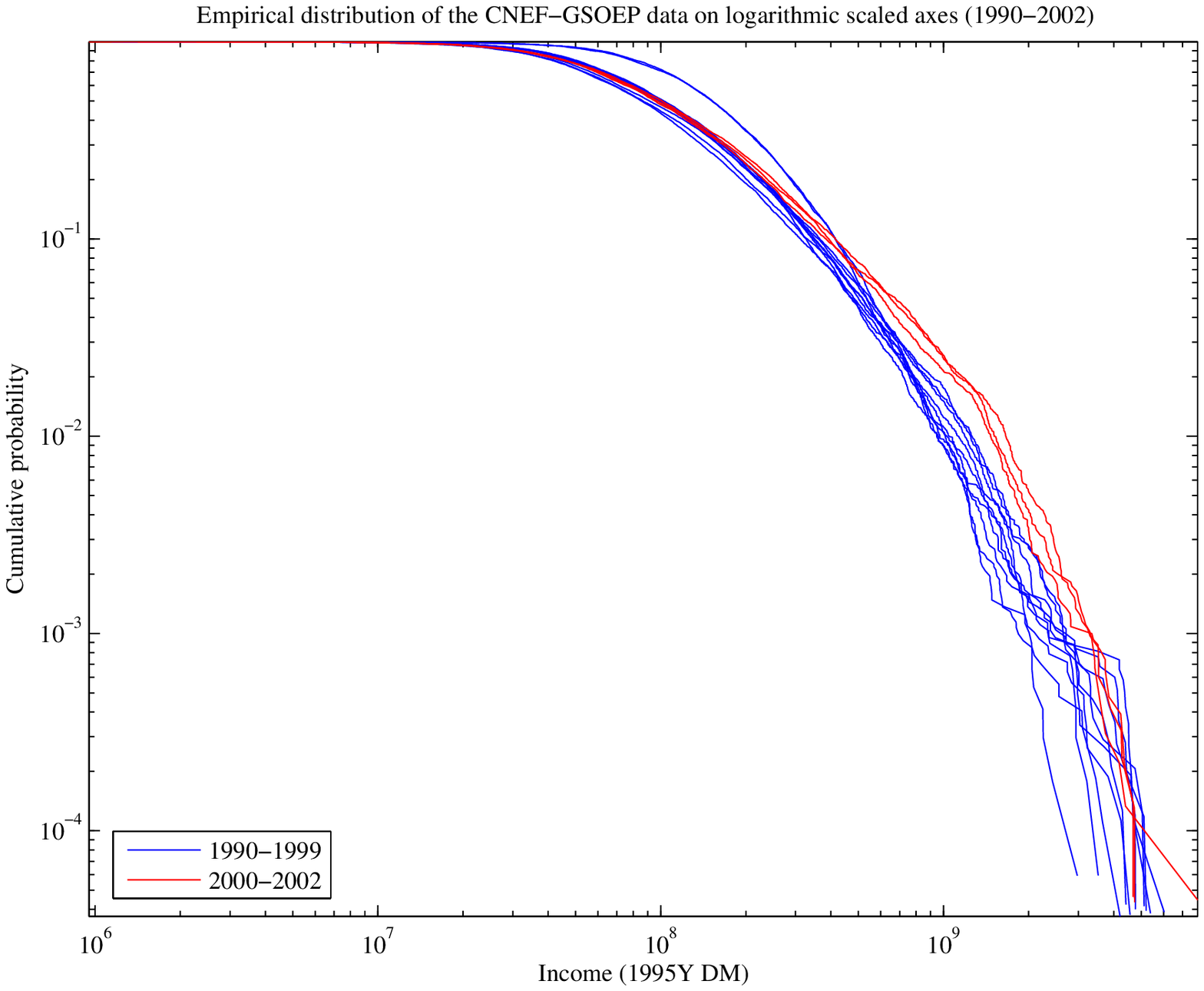}}
}
\caption{Time development of the income distribution for all the countries and years}
\label{fig:Temporal_Change_of_the_Distributions}
\end{center}
\end{figure}

\begin{figure}
\centering
\begin{center}
\mbox{
\subfigure[United States (1980--2001)]{\includegraphics[width=0.48\textwidth]{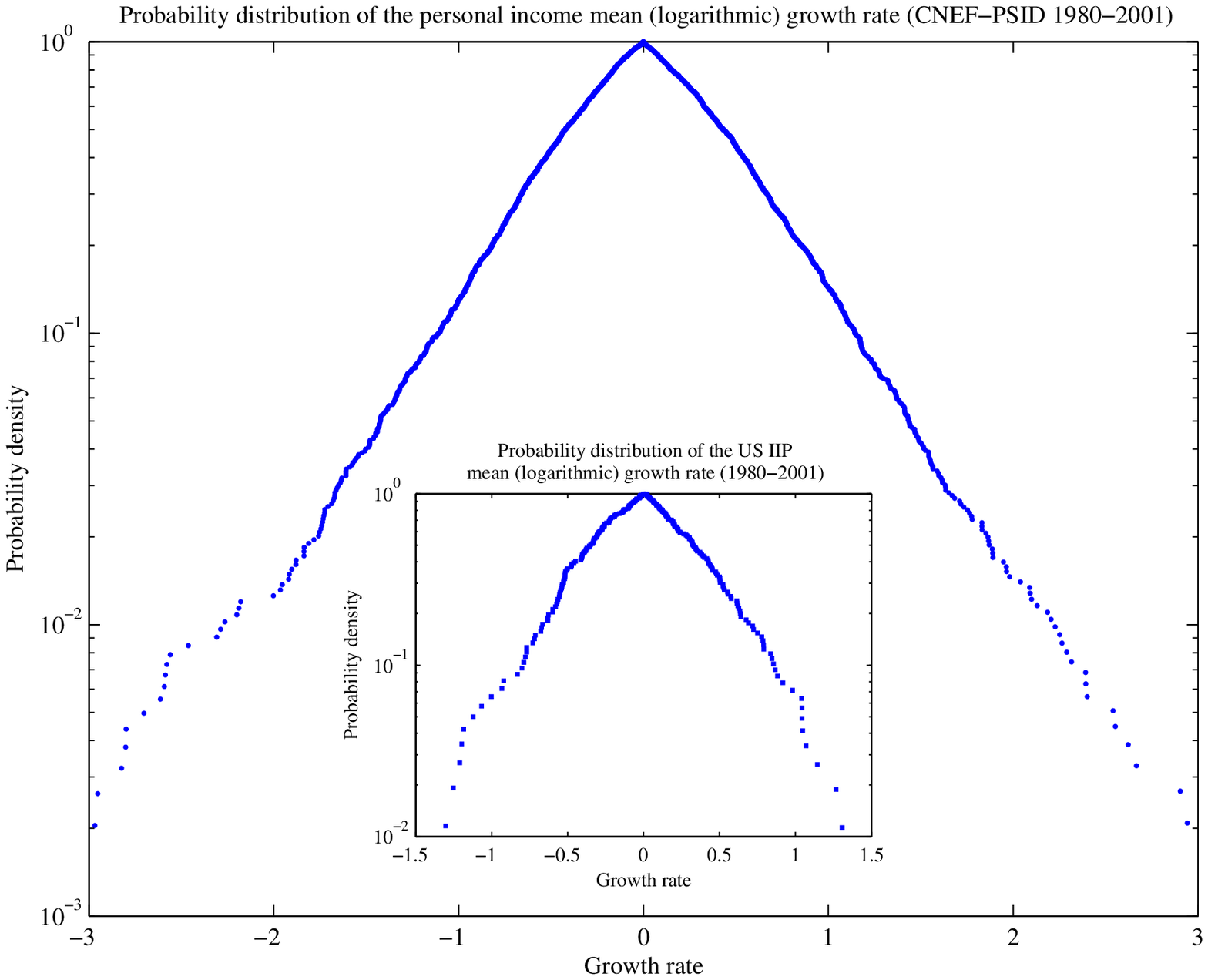}}
}
\mbox{
\subfigure[United Kingdom (1991--2001)]{\includegraphics[width=0.48\textwidth]{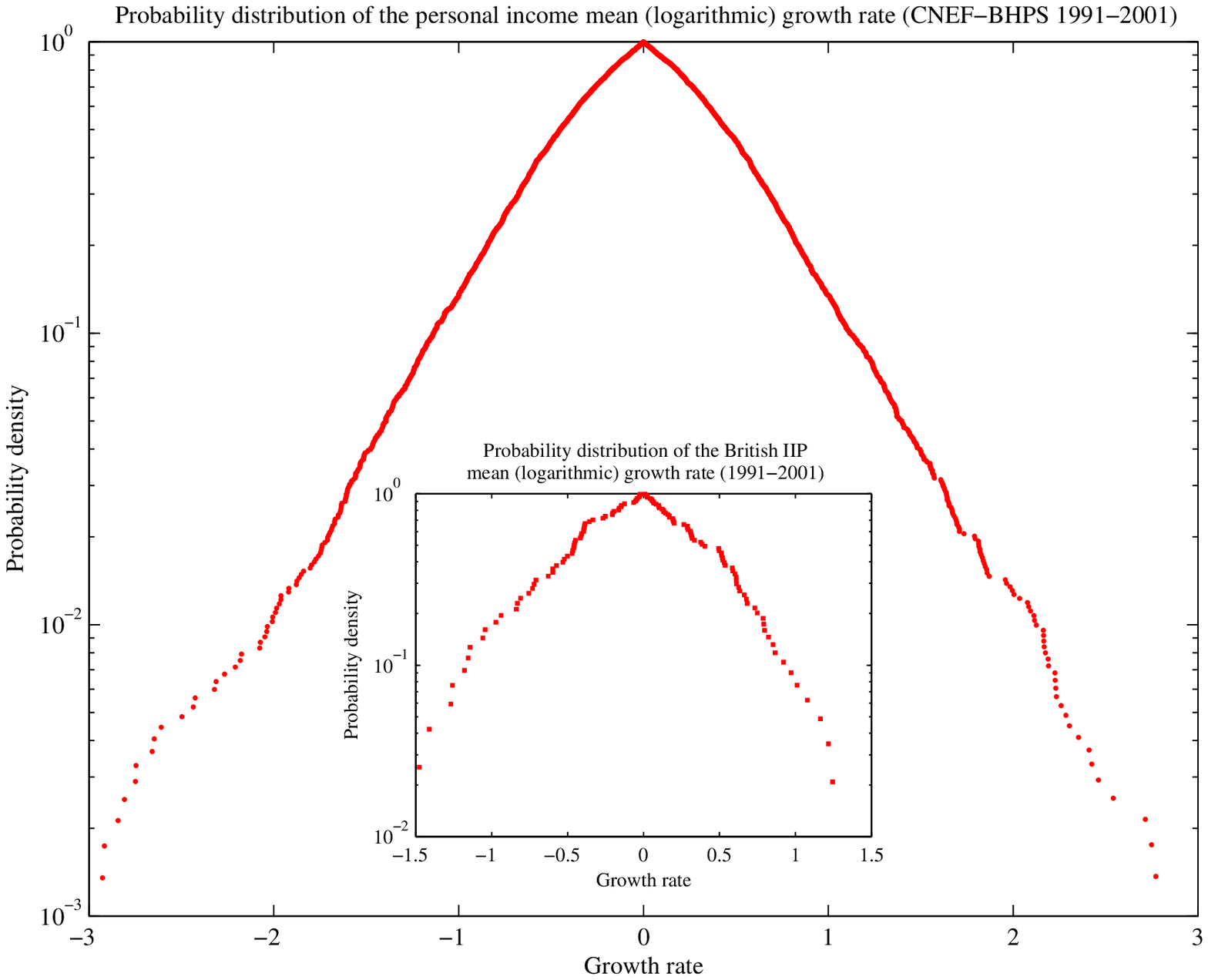}}
}
\mbox{
\subfigure[Germany (1990--2002)]{\includegraphics[width=0.48\textwidth]{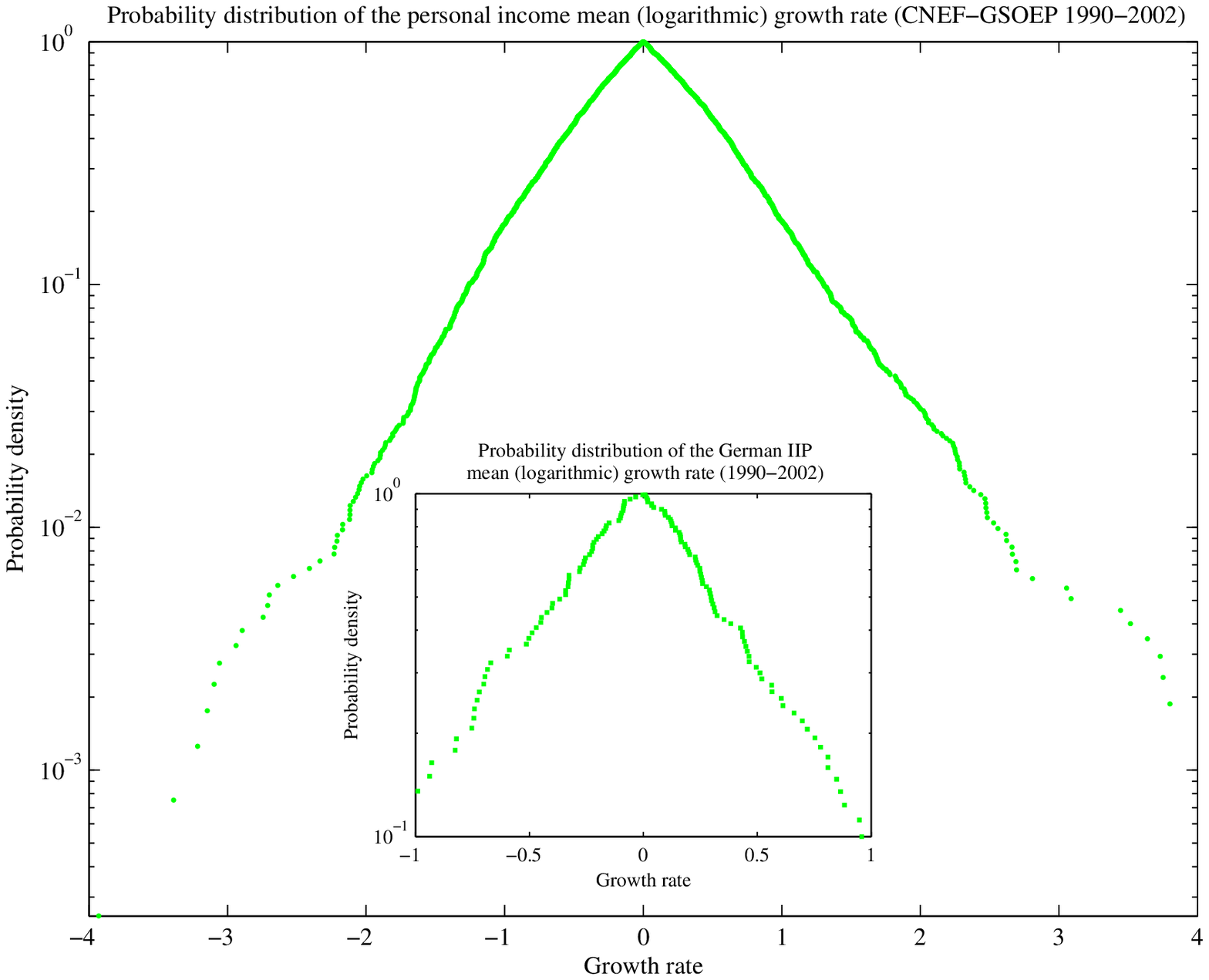}}
}
\caption{The probability distribution of equivalent income (main panels) and IIP (insets) growth rate for all the countries and years}
\label{fig:IIP_and_EI_Growth_Rates}
\end{center}
\end{figure}

The two-part structure of the empirical income distribution seems to hold all over the time span covered by our data sets. The distribution for all the years and countries are shown in Fig. \ref{fig:Temporal_Change_of_the_Distributions}.
As one can easily recognize, the distribution shifts over the years covered by our data sets. It is conceivable to assume that the origin of this shift consists in the growth of the countries. To confirm this assumption, we study the fluctuations in the output and equivalent income growth rate, and try to show that the evolution of both these quantities is governed by similar mechanisms, pointing in this way to the existence of a correlation between them as one would expect. We calculate the growth rates using the monthly series of the Index of Industrial Production (IIP) from [17] for output and connecting individual respondents' incomes over time for the equivalent income,\footnote{To properly weight the sample of individuals represented in all the years of the CNEF surveys, we use the individual's longitudinal sample weights.} and express them in terms of their logarithm.\footnote{All the data have been adjusted to 1995 prices and detrended by the average growth rate, so values for different years are comparable.} To account for the fact that the variance of the growth rates varies, we scale each growth rate by dividing by the corresponding estimated standard deviation. In Fig. \ref{fig:IIP_and_EI_Growth_Rates} we graph the empirical probability density function for these scaled growth rates, where the data points for the equivalent income in the main panels are the average over the entire period covered by the CNEF surveys.
As one can easily recognize, after scaling the resulting empirical probability density functions appear identical for observations drawn from different populations. Remarkably, both curves display a simple ``tent-shaped'' form; hence, the probability density functions are consistent with an exponential decay [18]:
\begin{equation}
f\left(r\right)=\frac{1}{\sigma\sqrt{2}}exp\left(-\frac{\left|r-\bar{r}\right|}{\sigma}\right)
\label{eq:Laplace_PDF}
\end{equation}
where $-\infty<r<\infty$, $-\infty<\bar{r}<\infty$, and $\sigma>0$. We test the hypothesis that the two growth rate distributions have the same continuous distribution by using the two-sample Kolmogorov-Smirnov (K-S) test; the results shown in Table \ref{K-S_Test} mean that the test is not significant at the 5\% level.
\begin{table}[h]
\centering
\caption{Two-sample Kolmogorov-Smirnov test statistics and {\em p}-values for both output and equivalent income growth rate data for all the countries}
\vspace{0.5cm}
\begin{tabular}{|c|c|c|}
\hline
\multicolumn{1}{|>{\columncolor{cLightYellow}}c}{Country}&\multicolumn{1}{|>{\columncolor{cLightYellow}}c}{K-S test statistic}&\multicolumn{1}{|>{\columncolor{cLightYellow}}c|}{{\em p}-value}\\
\hline
\rowcolor{cLightGray}United States&0.0761&0.1133\\
\hline
\rowcolor{cMediumGray}United Kingdom&0.0646&0.6464\\
\hline
\rowcolor{cLightGray}Germany&0.0865&0.2050\\
\hline
\end{tabular}
\label{K-S_Test}
\end{table}
These findings are in quantitative agreement with results reported on the growth of firms and countries [19--26], leading us to the conclusion that the data are consistent with the assumption that a common empirical law might describe the growth dynamics of both countries and individuals.
\par
\begin{figure}
\centering
\begin{center}
\mbox{
\subfigure[United States (1980--2001)]{\includegraphics[width=0.48\textwidth]{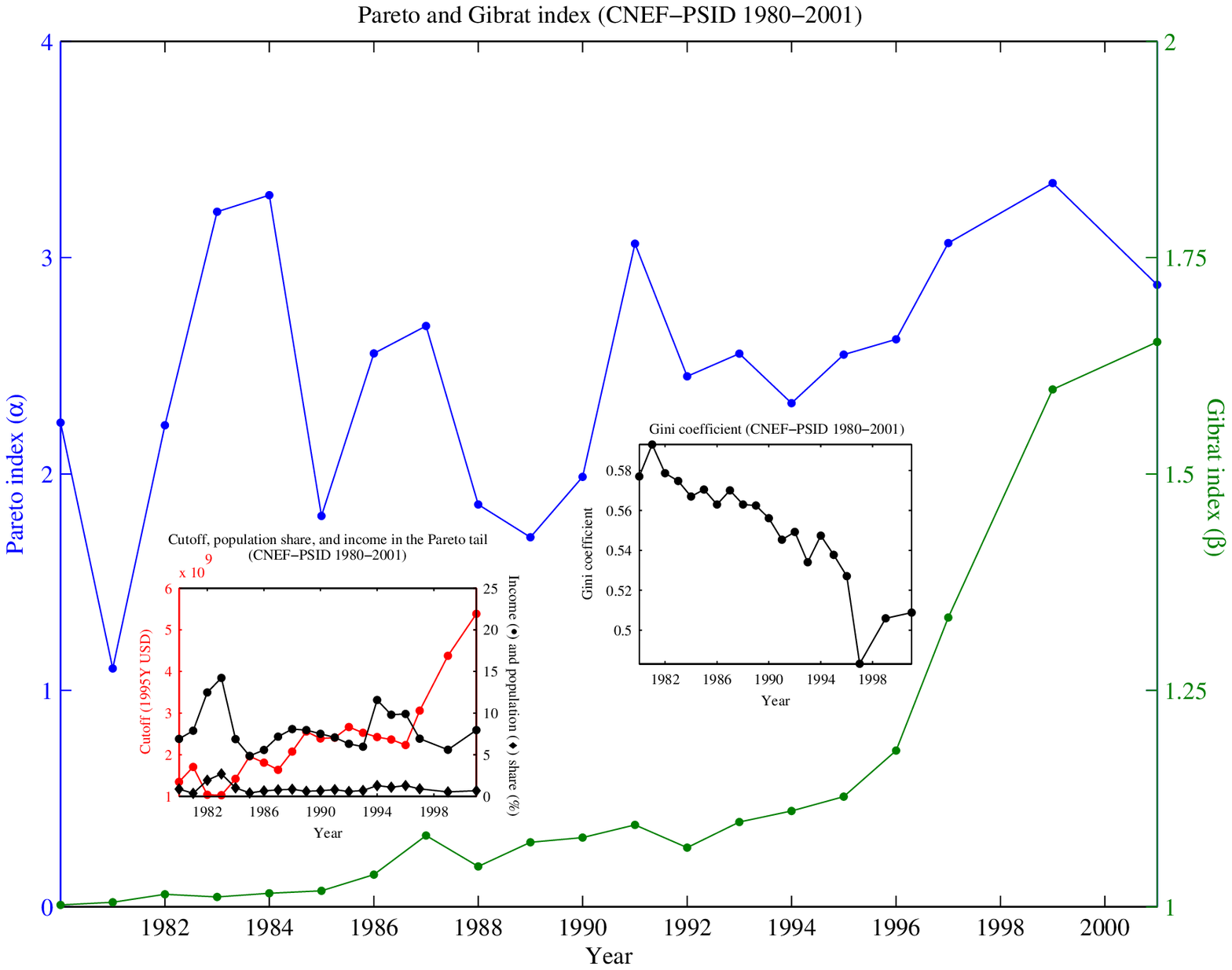}}
}
\mbox{
\subfigure[United Kingdom (1991--2001)]{\includegraphics[width=0.48\textwidth]{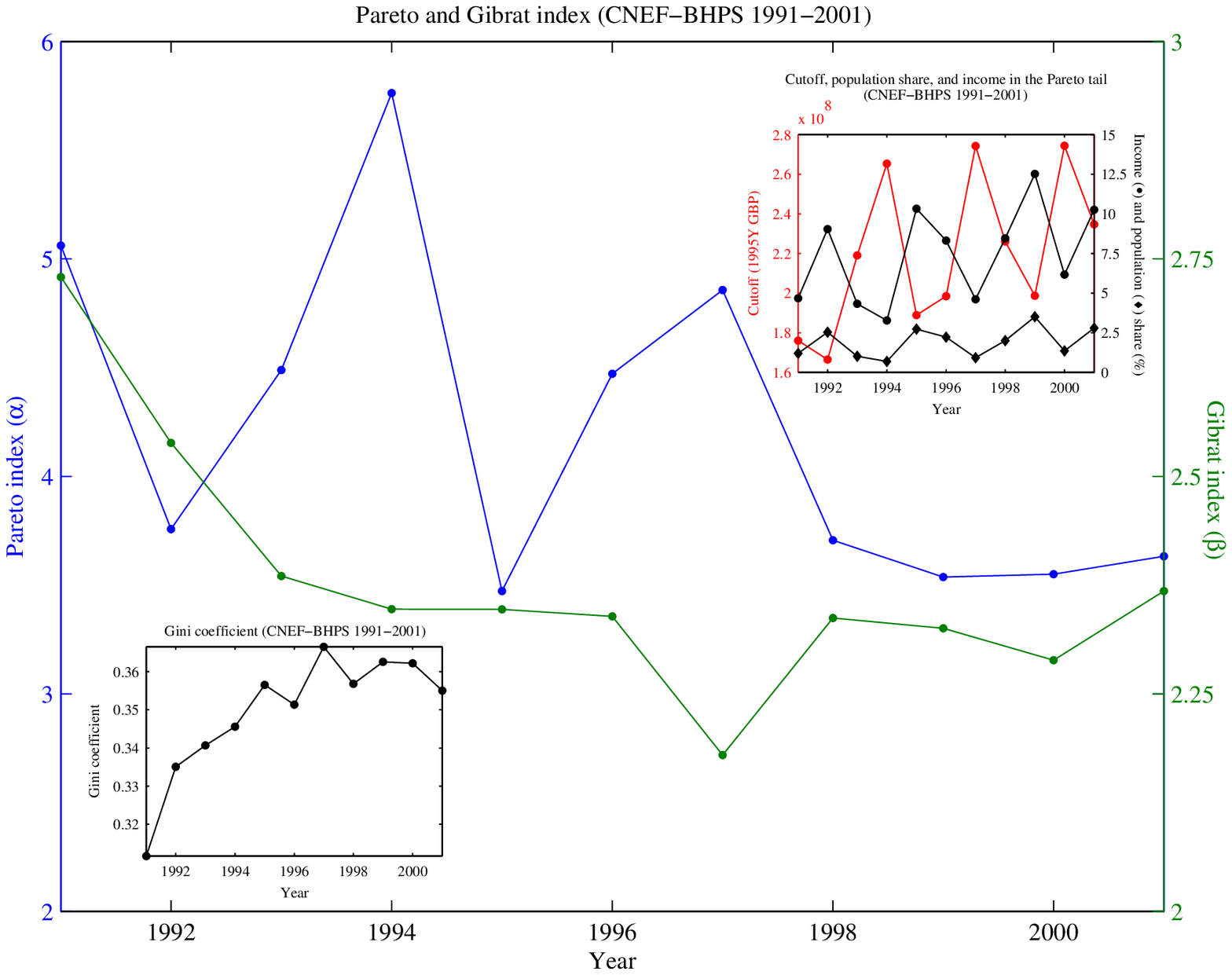}}
}
\mbox{
\subfigure[Germany (1990--2002)]{\includegraphics[width=0.48\textwidth]{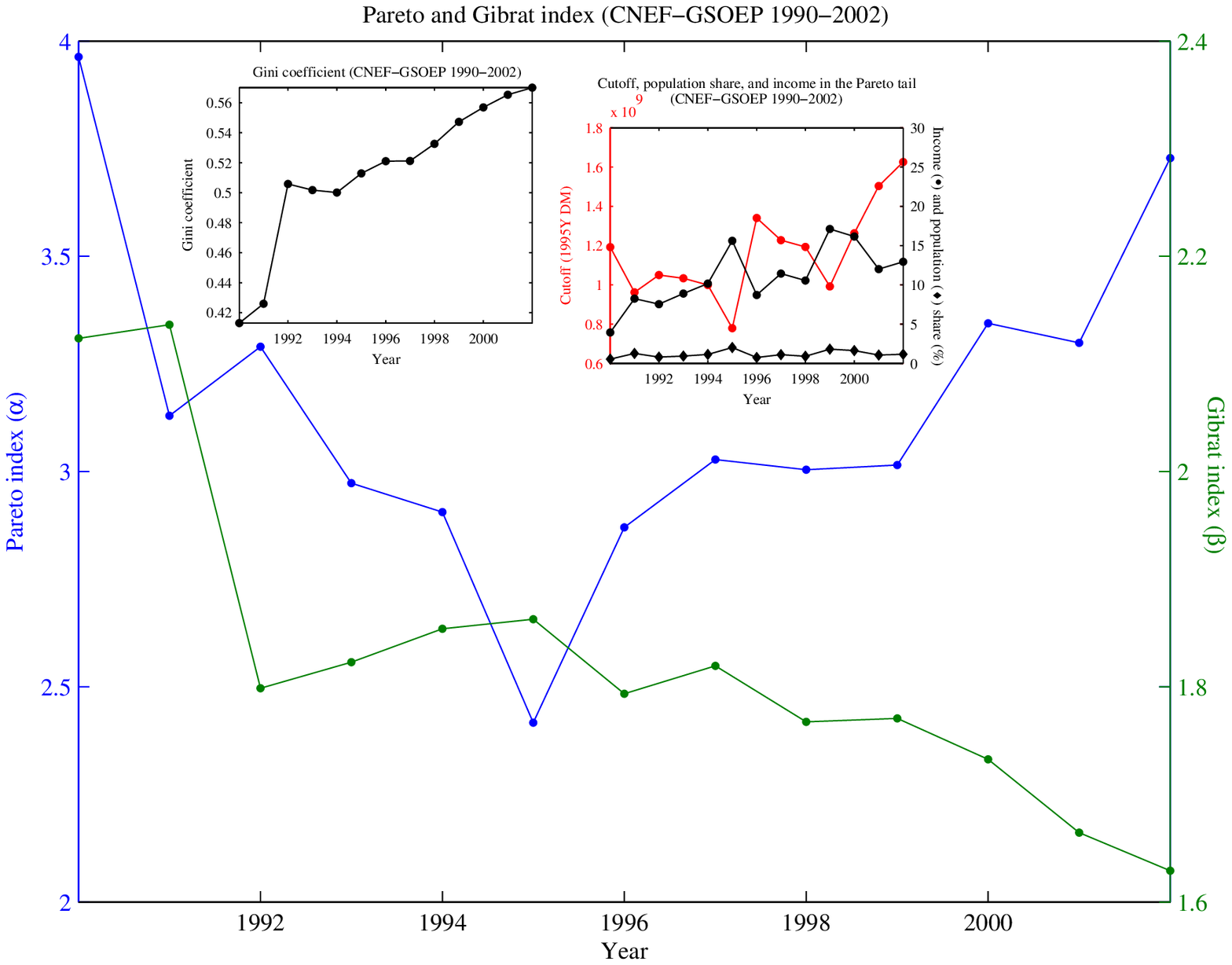}}
}
\caption{Temporal evolution of various parameters characterizing the income distribution}
\label{fig:Parameters}
\end{center}
\end{figure}

Even if the functional form of the income distribution expressed as lognormal with power law tail seems stable, its parameters fluctuate within narrow bounds over the years for the same country. For example, the power law slope has a value $\alpha=\left[1.1,3.34\right]$ for the US between 1980 and 2001, while the curvature of the lognormal fit, as measured by the Gibrat index $\beta=1/\left(\sigma\sqrt{2}\right)$, ranges between approximately $\beta=1$ and $\beta=1.65$; for the UK between 1991 and 2001, $\alpha=\left[3.47,5.76\right]$ and $\beta=\left[2.18,2.73\right]$; for Germany between 1990 and 2002, $\alpha=\left[2.42,3.96\right]$ and $\beta=\left[1.63,2.14\right]$. The time pattern of these parameters is shown by the main panels of Fig. \ref{fig:Parameters}, which also reports in one of the insets the temporal change of inequality as measured by the Gini coefficient.
As one can easily recognize, the information about inequality provided by the Gibrat index seems near enough to those provided by the Gini coefficient, which is a further confirmation of the fact that the lognormal law is a good model for the low-middle incomes of the distribution. The Pareto index is a rather strongly changing index. Among others, the definition of income we use in the context of our analysis contains asset flows. It is conceivable to assume that for the top 1\% to 3\% of the population returns on capital gains rather than labour earnings account for the majority share of the total income. This suggests that the stock market fluctuations might be an important factor behind the trend of income inequality among the richest, and that capital income plays an important role in determining the Pareto functional form of the observed empirical income distribution at the high income range [27]. The other insets of the pictures also show the time evolution of various parameters characterizing income distribution, such as the income separating the lognormal and Pareto regimes (selected as explained in Section \ref{sec:TheShapeOfTheDistribution}), the fraction of population in the upper tail of the distribution, and the share of total income which this fraction accounts for.\footnote{The share of total income in the tail of the distribution is calculated as $\mu_{\alpha}/\mu$, where $\mu_{\alpha}$ is the average income of the population in the Pareto tail and $\mu$ is the average income of the whole population.} One can observe that the fraction of population and the share of income in the Pareto tail move together in the opposite direction with respect to the cutoff value separating the body of the distribution from its tail, and the latter seems to track the temporal evolution of the Pareto index. This fact means that a decrease (increase) of the power law slope and the accompanying decrease (increase) of the threshold value $x_{R}$ imply a greater (smaller) fraction of the population in the tail and a greater (smaller) share of the total income which this population account for, as well as a greater (smaller) level of inequality among high income population.
\section{Summary and Conclusions}
\label{sec:SummaryAndConclusions}
Our analysis of the data for the US, the UK, and Germany shows that there are two regimes in the income distribution. For the low-middle class up to approximately 97\%--99\% of the total population the incomes are well described by a two-parameter lognormal distribution, while the incomes of the top 1\%--3\% are described by a power law (Pareto) distribution.
\par
This structure has been observed in our analysis for different years. However, the distribution shows a rightward shift in time. Therefore, we analyze the output and individual income growth rate distribution from which we observe that, after scaling, the resulting empirical probability density functions appear similar for observations coming from different populations. This effect, which is statistically tested by means of a two-sample Kolmogorov-Smirnov test, raises the intriguing possibility that a common mechanism might characterize the growth dynamics of both output and individual income, pointing in this way to the existence of a correlation between these quantities. Furthermore, from the analysis of the temporal change of the parameters specifying the distribution, we find that these quantities do not necessarily correlate to each other. This means that different mechanisms are working in the distribution of the low-middle income range and that of the high income range. Since earnings from financial or other assets play an important role in the high income section of the distribution, one possible origin of this behaviour might be the change of the asset price, which mainly affects the level of inequality at the very top of the income distribution and is likely to be responsible for the power law nature of high incomes.
\begin{ack}
The authors wish to thank the organizers of the \emph{International Workshop on} \texttt{Econophysics of Wealth Distributions} (ECONOPHYS - KOLKATA I, 15--19 March, 2005, SINP, Kolkata) for the very nice hospitality and all the invited participants for the useful interactions.
\end{ack}

\end{document}